\documentclass{article}
\usepackage{graphics}
\usepackage[xdvi]{epsfig}
\newcommand{\emaila}{e.fischer.stolberg@t-online.de}
\begin{document}
%\shorttitle{Does gravitational collapse lead to singularities?}
%\shortauthors{Ernst Fischer}
\title{Does gravitational collapse lead to singularities?}
\author{Ernst Fischer\footnote{\emaila}}
\maketitle

%\keywords{cosmology; black hole}

\begin{abstract}
According to conventional modelling by general relativity the
collapse of radially symmetric gravitating objects may end in a
singular state. But by inclusion of potential
energy into the energy tensor, which is required to guarantee global
energy conservation, the occurrence of singularities is avoided.
Instead the final states of the collapse of mass concentrations of
arbitrary size are nuclear matter objects, from which jets of
matter can be recycled into space. The mysterious dark energy,
supposed as the main constituent of the universe, may even be the
potential energy of matter itself.
\end{abstract}

\section{Introduction}
Einstein's theory of general relativity (GRT) is regarded today as
the best description of gravitational interaction. The concept to
interpret motion under the influence of gravitation as geodesic
motion in a non-Euclidean space-time continuum has successfully
passed all observational tests. The aberration of light passing
close to the sun, the perihelion shift of Mercury or other planets
and the frequency shift of light by gravitational interaction have
impressively confirmed the theoretical model.

But all these observations only test the range of weak
gravitational fields. That means that interactions of test bodies
or quanta can be described by geodesic motions in a geometry,
which is set up by a given matter distribution. In the strong
field range, where every element of matter contributes to geometry
as well as it is object of changes, observations are less
conclusive. Though the gravitational collapse of stars or galaxies
into objects of extreme density appears as observationally
confirmed, the details of these processes are far from being well
understood. According to the textbooks of general relativity (see
e.g. Wald \cite{wald} or Hawking \& Ellis \cite{hawking}) massive objects may contract
into a final state, from which an escape of matter is impossible,
and finally even into a singularity, a state of infinite matter density. But to all our experience in other parts of physics singularities do not exist in nature as physical entities. The occurrence of singularities in the mathematical description is always the consequence of an inaccurate modelling or unallowed extrapolations. Thus we must rise the question, if the singular states occurring in GRT should be regarded as a purely mathematical approximation. There is no observational evidence that the collapse to a singular state really happens. 

Thus it appears reasonable to look somewhat closer to the physics
of the collapsed objects, normally denoted as black holes. Is it
possible to understand their physics within the geometrical
concept of general relativity, but without creating mathematical
singularities? In this article we will show that this is possible,
if we only apply the concept of energy conservation more strictly
than in conventional models. We will concentrate on the description
of spherically symmetric objects in static equilibrium, as the correct 
description of these equilibria must be regarded also as the basis 
of all dynamical developments.

\section{Spherically symmetric solutions}
Spherically symmetric objects in GRT are defined by the condition 
that in the field equation
\begin {equation}
\label{field} R_{ij}-\frac{1}{2}R\, g_{ij}=\kappa T_{ij}
\end{equation}
as well the Ricci tensor $R_{ij}$, derived from the metric $g_{ij}$, 
as the elements of the energy tensor $T_{ij}$, which contains the sources of gravity, depend only on one spatial parameter. This radial parameter can be defined in different ways.
In Euclidean geometry it is normally identified with the length $r$ of the shortest line  
connecting some point with the center of symmetry. This is 
identical with the definition that $r$ is the length of 
a closed line at constant parameter $r$, divided by $2\pi$.

If space is curved, these two definitions do no longer agree. 
If we write the line element in the form
\begin {equation}
\label{ds} ds^2=-f(r,t)dt^2+h(r,t)dr^2+r^2d\Omega^2
\end{equation}
where $t$ is the time parameter, $r$ a radial parameter and
$d\Omega^2=d\vartheta^2+cos^2\vartheta d\varphi^2$ defines the
element of solid angle $d\Omega$, $r$ is defined by the length of a circle divided by $2\pi$.
This fact we should keep in mind, when we try to do calculations in 
curved space. 

One well known example of a radially symmetric problem is the so called
vacuum solution of the field equations, first derived by Schwarzschild 
\cite{schwarz}, which describes a system, in which
the geometry is dominated by a central gravitating object, so that
the influence of masses outside of this object on the metric can be
neglected. In this case the field equations are reduced to the
vanishing of the Ricci tensor $R_{ij}=0$. Of course, the
denomination 'vacuum solution' is somewhat misleading, as normally it is not used to
describe a vacuum, but to describe motions or interactions
of matter outside the central object, only that the influence of this matter or
radiation on the geometry is negligible compared to
that of the central gravitating object.

Under these conditions the field equations are reduced to
\begin{equation}
\label{hh}
\frac{1}{hr^2}\frac{dh}{dr}+\frac{1}{r^2}\left(1-\frac{1}{h}\right)=0,
\qquad \frac{1}{f}\frac{df}{dr}=-\frac{1}{h}\frac{dh}{dr}
\end{equation}
with the solution $f(r)=1/h(r)=1-C/r$. To derive this formula use has been made 
of the special form of the coordinate system, in which the angular part of the 
metric is defined by the condition that the length of a closed line with $r=const.$ is $2\pi r$.
From comparison of the weak field limit with Newtonian gravity then
the parameter $r$ is identical with the radial
coordinate and the constant $C$ is related to the gravitating mass at
the center. This leads to the formula
\begin {equation}
f(r)=1-\frac{r_s}{r} \qquad
h(r)=\left(1-\frac{r_s}{r}\right)^{-1}
\end{equation}
where $r_s$ is the well known Schwarzschild radius $r_s=2GM/c^2$
($G$ is the gravitational constant and $M$ the mass of the central
object). But already with this simple solution we get into conceptual
problems, if we try to extend it to processes which take place in regions close to the Schwarzschild
radius. Assuming that space outside $r_s$ contains matter of density
$\varrho(r)$, the amount of matter in a spherical shell of thickness $dr$ at $r=r_s$, expressed by our coordinates, is given by
\begin {equation}
\label{mass}
dm=\sqrt{h}\varrho(r)r^2dr\,d\Omega=4\pi\frac{\varrho(r)r^2}{\sqrt{1-r_s/r}}dr
\end {equation}
But even if the density is arbitrarily small but not exactly zero
at $r=r_s$, this yields an infinite value, in contradiction to the
assumption that the mass is negligibly small. For $r<r_s$ a real value of $dm$ does not even exist. This singular
behavior is caused by the fact that the size of the spatial volume
element $dV=4\pi r^2 \sqrt{h(r)} dr$ becomes infinite at $r=r_s$.

There are two possible ways out of this dilemma. Either the
definition of the radial parameter is no longer valid in the strong field regime or the
definition of energy or matter density  breaks down and must be
modified. But the definition of the radial parameter leaves no
room for a recalibration, as on one hand in the limiting case of
weak gravitation it should agree with the radial distance and on
the other hand by the definition of the angular part of the line
element by eq.(\ref{ds}) the parameter $r$ is uniquely defined. There is no 
physical reason, why a closed line with length $s\leq 2 \pi r_s$ should not exist. 
So the only way out is that the parameter $\varrho$ is not adequate to describe 
the matter or energy distribution and we have to redefine the density parameter.

Anyway a unique definition of densities appears as a problem in a
system, in which the size of the spatial volume element varies,
when the amount of matter in the neighborhood changes. If we
propose that matter consists of individual particles with their
rest mass as an invariant property and that the number of those
particles in some volume, which is defined by fixed limits, cannot
change, when the matter distribution outside this volume is
changed, local densities, defined as the amount of some quantity per volume, cannot be regarded as invariant
properties of the spatial distribution.

The proposition of Einstein, to use the densities as defined in the local
Euclidean tangent space, can be regarded as a suitable approximation in the
weak field regime, but it appears invalid in strong fields. Incorporating
gravitation into the geometry of space requires incorporating curvature into
the definition of density as well. The matter density as defined in the Euclidean
tangent space must be regarded as a local parameter. But if we want to determine 
the total matter content in some volume, defined by limiting values of the radial parameter,
we cannot simply do it by summing up the locally defined tangent space densities, that means, by integration of $dm$ as given by eq.(\ref{mass}). To define a conserved 
integral property like the total mass, we must add a correction term to the integrand, 
accounting for the change of volume by curvature. In the definition of density we have to replace the size of the 
tangent space volume element by the corresponding element in curved space. The volume element $dV_0=4\pi r^2dr$ has 
to be replaced by $dV=4\pi r^2/\sqrt{h}dr$.

What this means becomes clear immediately, considering again
the weak field limit. According to the definition of the line element eq.(\ref{ds})
the mass in a spherical shell between radii $r_1$ and $r_2$ with density $\varrho$ is
\begin {equation} \Delta m=4\pi\int_{r_1}^{r_2}\sqrt{h(r)}\varrho(r)r^2dr=4\pi
\int_{r_1}^{r_2}\frac{\varrho(r)r^2}{\sqrt{1-r_s/r}}\,dr.
\end {equation}
With $r_1,r_2\gg r_s$ it can be approximated by
\begin{equation}
\Delta m=4\pi\int_{r_1}^{r_2}\varrho(r)r^2dr+4\pi\int_{r_1}^{r_2}\frac{GM\varrho(r)r}{c^2}\,dr.
\end{equation}
The second term is just the mass equivalent of the gravitational
binding energy or of the negative of the potential energy of the
system. Integration of the density, as defined in Euclidean
tangent space results in an overestimate of the total mass. Thus
it appears logical to consider the summed density of mass energy
and potential energy as the equivalent to mass energy in Euclidean
space. With this redefinition at the Schwarzschild radius the
matter energy density $\varrho c^2$, defined in the local tangent space, 
just balances its own potential energy $\varrho c^2 (\sqrt{1-r_s/r}-1)$, so that the
effective contribution to the total energy remains negligibly small.

Including potential energy into the energy balance in this way, a
basic property of Newtonian gravity is recovered, the conservation
of energy. Without inclusion of potential energy into the balance,
gravitational collapse would be accompanied by a continuous gain
of energy from the gravitational field. But in the geometrical
concept of general relativity there exists no gravitational field
which might possess energy. Gravitation is only a consequence of curvature.
So this energy is created from nothing. Only if we include potential 
energy of matter itself into the balance, conservation of energy is guaranteed 
also in systems of strong curvature.

The question remains, how to define a general expression of
potential energy within the formalism of GRT. It is a basic
concept of GRT that it is a strictly local theory. That means that
the complete information, which describes the behavior of matter
under the influence of gravitation is, besides of locally defined
parameters, contained in the metric tensor.

That Einstein used the field equation (\ref{field}) instead of simply 
setting $R_{ij}=T_{ij}$ was motivated by the fact that the divergence 
of the left hand side should be zero and thus represents a conserved quantity, 
as he took for sure that the energy tensor on the right hand side was a 
conserved quantity, too. But the fact that the tensor divergence vanishes locally is not sufficient to guarantee conservation also on integral scale. Exact conservation of the total energy is possible only, if we include 
potential energy into the definition of the energy tensor.

In the case of a radially symmetric solution the only reasonable way to describe
potential energy is a term in the energy tensor of the form $\lambda(r) g_{ij}$, where the scalar function $\lambda$ depends only on local parameters, on the tangent space matter density
$\varrho$ and the local pressure $P$ (leaving out of consideration
radiation and other possible minor contributions to the energy field for simplicity).

In curved space we must distinguish between two types of local
parameters, those which represent intensive properties of
particles resp. their local mean and densities of extensive
properties, defined as the amount of some property per volume.
When, as in general relativity, geometry depends on the matter
distribution, the latter group of parameters must be adjusted for
the variability of volume, if we want to make a transition to
macroscopic quantities by integration.

A typical quantity of this latter kind is the particle density. If
a fixed number of particles is contained in some volume defined by
its limits, this number will not change, when the geometry changes
by adding matter somewhere outside this volume. The same holds for
the matter content, if we assume that every particle has a
conserved property, its rest mass. Thus the effective matter
density must change with the geometry of space.

To determine macroscopic data like the total mass of a body from
local quantities by integration, we have to replace the local
matter density, as defined in Euclidean tangent space, by the sum
of this density and the matter equivalent of its potential energy.
We have to replace the quantity $\varrho$ by $\varrho /
\sqrt{h(r)}$ , where $h(r)$ is the quantity defined by
eq.(\ref{ds}), describing the deviation from Euclidean geometry.

\section{Interior solutions}
By now we have only addressed the problems occurring outside the 
Schwarzschild radius. But the question put in the beginning was, 
what happens in the interior of a matter distribution in the case 
of strong gravitation. Inclusion of potential energy into the balance equations does not
only influence the exterior of the Schwarzschild solution, but
also the balance in the interior of any spherically symmetric matter
distribution.  The field equation for the static spherically symmetric case is well known as the 
Tolman-Oppenheimer-Volkoff (TOV) equation \cite{TOV} 
\begin{equation}
\label{tov} \frac{dP}{dr}=-(\varrho+P)\frac{m+4\pi r^3P}{r^2(1-2m/r)},
\end{equation}
(in units, where G=c=1), which relates the local equilibrium pressure $P$ to the matter
distribution. In the conventional description the quantity $m(r)$ results from integration of the
density $\varrho$ over a sphere of radius $r$.
\begin{equation}
\label{m}
m(r)=4\pi \int_0^r \varrho r^2 dr.
\end{equation}
The component $h(r)$ of the metric tensor is related to the density by
\begin{equation}
\label{hdiff}
\frac{1}{rh^2}\frac{dh}{dr}+\frac{1}{r^2}\left(1-\frac{1}{h}\right)=8\pi \varrho
\end{equation}
which is connected to $m(r)$ by $h(r)=1/(1-2m(r)/r)$.

But when space is curved and we want to determine the influence of matter inside a sphere of radius $r$ on the metric by integration of eq.(\ref{m}), inserting the density as defined in 
Euclidean tangent space will not give the correct result. As has been discussed in the last section, we have to include the potential energy term into this equation, replacing the Euclidean matter 
density $\varrho$ by $\varrho+\lambda(r)$ with
\begin{equation}
\lambda(r)=\varrho(1/\sqrt{h}-1)
\end{equation}

Thus the integrand of eq.(\ref{m}) implicitly depends on the
metric and by this on $m(r)$, so that the function $m(r)$ now has
to be determined from the differential equation
\begin{equation}
\frac{dm}{dr}=4\pi r^2\varrho\sqrt{1-2m/r}
\end{equation}
with the boundary condition $m(0)=0$. This leads to a completely
different form of the solution.

This can be best demonstrated assuming as an example a system of
constant density $\varrho_0$ and radial extension $R_0$. It is
reasonable to assume that in every real isolated matter
distribution the condition $d\varrho/dr<0$ holds, so that with
$\varrho=\varrho_0$ the maximum deviation of $h(r)$ from unity
will be obtained. In this case introducing a normalized coordinate
$x=r/R$ with $8\pi\varrho_0 R^2=1$ and the new variable $y=x/h$
eq.(\ref{hdiff}) can be written in the form
\begin{equation}
\label{ydiff} \frac{dy}{dx}=1-x^{3/2}y^{1/2},
\end{equation}
while without potential energy we have
\begin{equation}
\frac{dy}{dx}=1-x^2
\end{equation}
Fig.1. shows the resulting function $h(x)$ for both equations.
Without potential energy the solution exhibits a singularity at
$x=\sqrt{3}$, while with potential energy $h(x)$ is finite for all $x$.
\begin{figure}[hbt]\epsfig{file=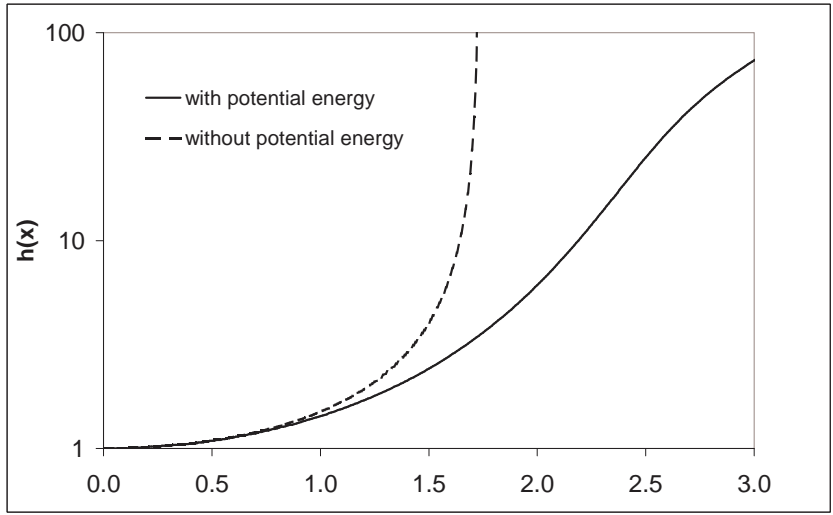,angle=-90,width=11cm}
\caption{Radial metric parameter $h(x)$ as a function of the
normalized coordinate $x=r/\sqrt{8\pi\varrho_0}$}
\end{figure}
The solution of eq.(\ref{ydiff}) with $y(0)=0$ exhibits an extremum near $x=1.2$ and then decreases monotonically towards zero with increasing x. That means that the quantity $2m(x)/x=1-1/h(x)$ can never reach the value one for any reasonable matter distribution. This value of $2m/x$ must be regarded as an upper bound also for other density distributions. With other words: The Schwarzschild limit, which requires a value of one, can never be reached for any reasonable matter distribution, if only the condition $d\varrho/dr<0$ is satisfied. The Schwarzschild radius must be regarded as a purely mathematical quantity. Under the assumption of a monotonic density profile a central object of
arbitrary mass cannot exist in a volume limited by $r_s$.

The fact that the quantity $(1-2m(r)/r)$ can never reach zero is
essential also for the solution of the pressure balance
eq.(\ref{tov}). The denominator of the TOV equation is always
positive. No infinite pressure is ever necessary to balance
gravitational force. Due to the form of the potential energy term
$\lambda(r)\,g_{ij}$ there is also a negative contribution of
curvature to the pressure, but this does not change the general
form of the solution.

The solution of the system of eqs.(\ref{tov}) and
(\ref{m}) requires the knowledge of an equation of state, some
functional relation between $\varrho$ and $P$, just like in
conventional modelling of stars, where potential energy is
neglected. But there is no situation, in which a static balance is
impossible so that collapse would proceed into a singular state. The
only proposition for a stable equilibrium configuration is that
$\varrho$ and $P$ decrease monotonically from the center to the
surface of the matter distribution.

In normal stars, where gravitational attraction is balanced by
thermal pressure produced by thermonuclear reactions, pressure and
density are always monotonic functions of the radial coordinate
and besides that, the influence of potential energy on the balance
is negligibly small. This remains true also when the thermonuclear
fuel is consumed and the star cools off by radiation. In this case
the degeneracy pressure of electrons or neutrons takes over, to
stabilize the star against further gravitational collapse. There
may occur unstable situations, as e.g. in white dwarfs, where the
degeneracy pressure decreases with increasing density, when
electrons and protons recombine into neutrons. This leads to
instabilities and stellar explosions, as we know them as
supernovas. But finally every collapsing object can end in an
equilibrium state, where degeneracy pressure of the constituting
particles balances gravitation. There is no upper mass limit
beyond which infinite pressure would be required. The occurrence of an horizon and of singularities in GRT is only the consequence of an improper definition of the energy tensor.

\section{Discussion}
In the last section we have demonstrated that an equilibrium state of collapsed
matter will always be outside the singular conditions assumed in
the Schwarzschild geometry. There is no horizon, from which no escape of matter or radiation is possible and black holes as singular points of infinite density cannot exist. Instead every spherically symmetric
gravitational collapse can find a final equilibrium state of
finite density. No matter is inevitably lost from the surrounding
space.

Of course, by now we have only discussed equilibrium
configurations. In nature such idealized equilibria are scarcely
reached. Most collapsing objects continuously accrete matter from
surrounding space. Besides in most cases there is no spherical symmetry. Accreting objects
will accumulate some angular momentum and the inflow of matter may be
concentrated to the rotational plane. Magnetic fields can also
influence the balance, if the particles are electrically charged.

Thus during the formation of collapsing objects unstable
situations may occur, leading to expulsion of the outer matter
shell or to the nearly complete disruption of a star. But in a
final state, when degeneracy pressure dominates the dynamics,
kinetic pressure of incoming matter is negligible. It is only
rotation of the complete system, which may influence the dynamical
equilibrium. By principle rotation velocities may be close to the
velocity of light, so that the rotational energy is comparable to
the rest energy of particles, just like the degeneracy energy. In
this case pressure is no longer isotropic. Inflow of matter near
the rotational plane will cause an outflow along the axis of
rotation. The observed formation of cosmic matter jets from
collapsed stars or active galaxy cores can be understood more
easily, if we take as a fact that the 'black hole' in the center is not a
matter concentration on the other side of some semi-permeable
horizon, but an accumulation of nuclear matter, which can be
recycled into the universe under suitable conditions. This is of
essential importance to understand the physics of the matter jets
emerging from the supermassive 'black holes' in the cores of
active galaxies. Detailed modelling of these cores is still
missing. But to understand them, the first requirement is to
start from the correct balance equations.

Neglecting potential energy in the balance equations appears as a
general problem in the conventional methods of general relativistic
modelling. In the description of the global dynamics of the
universe, in addition to the search for dark matter, people are looking for the so called dark energy, which
is necessary to bring the theoretical model into agreement with
observations. This dark energy should be present throughout the
universe and exhibit a negative pressure and an energy density
comparable in order of magnitude to that of matter. Potential
energy of matter itself just fulfills all these requirements. In a
homogeneous solution of the Einstein equation it would look just
like a cosmological constant, with the only difference that it is
not a true constant, but varies with the matter density. No
mysterious dark energy is necessary to fulfil the balance. Potential
energy of matter itself can do the job.

It should be mentioned in this context that with this additional
term instead of a cosmological constant the static universe
proposed by Einstein would be stable, as with varying radius $a$
of the universe the potential energy varies as $\varrho/a$, so
that a virtual increase of $a$ would produce a negative $da/dt$.
The entire universe would be the only system which exists at its Schwarzschild
radius, the state at which the total matter energy is balanced by
its own potential energy. 

By now we have no secure confirmation that general relativity
delivers the correct description of the universe, but with the
corrections discussed in this paper some of its shortcomings, the
existence of singularities and the missing explanation of the mysterious dark
energy are automatically resolved. Also the order of magnitude agreement between dark energy and matter density
can be understood. The only thing we have to do is, to
accept the principle of energy conservation not only locally but
also on macroscopic and global scale, independent of the geometry
of space.

\end{document}